\documentclass[12pt]{article}
\usepackage{amsmath,amssymb,amsthm,amsxtra,overpic,bbm,bm,epsfig}
\def\thefootnote{\fnsymbol{footnote}}

\begin{document}

\vspace{0.2cm}

\begin{center}
{\Large\bf A U-spin prediction for the CP-forbidden transition
$e^+ e^- \to D^0\bar{D}^0 \to \left(K^+ K^-\right)^{}_D
\left(\pi^+ \pi^-\right)^{}_D$}
\end{center}

\vspace{0.2cm}

\begin{center}
{\bf Zhi-zhong Xing$^{a, b}$}\footnote{Email: xingzz@ihep.ac.cn} \\
{$^a$Institute of High Energy Physics, and School of Physical
Sciences, \\ University of Chinese Academy of Sciences, Beijing 100049, China \\
$^b$Center for High Energy Physics, Peking University, Beijing
100080, China}
\end{center}

\vspace{1.5cm}

\begin{abstract}
The LHCb Collaboration has recently reported the discovery of direct CP violation in
combined $D^0 \to K^+ K^-$ and $D^0 \to \pi^+ \pi^-$ decay modes at the $5.3 \sigma$
level. Assuming U-spin symmetry (i.e., $d \leftrightarrow s$ interchange symmetry)
for the strong-interaction parts of these two channels, we find that their corresponding
direct CP-violating asymmetries are
${\cal A}^{}_{\rm CP} (K^+ K^-) \simeq \left(-7.7 \pm 1.5\right) \times 10^{-4}$ and
${\cal A}^{}_{\rm CP} (\pi^+ \pi^-) \simeq \left(7.7 \pm 1.5\right) \times 10^{-4}$.
The CP-forbidden transition $e^+ e^- \to D^0\bar{D}^0 \to \left(K^+ K^-\right)^{}_D
\left(\pi^+ \pi^-\right)^{}_D$ on the $\psi(3770)$ resonance is therefore expected
to have a branching fraction of ${\cal O}(10^{-10})$ or smaller under U-spin symmetry, 
and it can be observed at a high-luminosity super-$\tau$-charm factory if at least 
$10^{10}$ pairs of coherent $D^0$ and $\bar{D}^0$ events are accumulated.
\end{abstract}

\begin{flushleft}
\hspace{0.8cm} PACS number(s): 14.60.Pq, 11.30.Hv, 13.35.Hb.
\end{flushleft}

\newpage

\def\thefootnote{\arabic{footnote}}
\setcounter{footnote}{0}

\framebox{\bf 1} ~ Within the standard model {\it charmed} CP violation
in $D$-meson decays is expected to be of ${\cal O}(10^{-3})$ or
smaller. The reason for this expectation is simply that the
{\it charmed} unitarity triangle of the Cabibbo-Kobayashi-Maskawa
(CKM) quark flavor mixing matrix $V$, defined by the orthogonality
relation $V^*_{ud} V^{}_{cd} + V^*_{us} V^{}_{cs} + V^*_{ub}
V^{}_{cb} = 0$ in the complex plane as illustrated by Fig. 1 \cite{FX00},
is so sharp that the ratio of the CP-violating part to the
CP-conserving part in many $D$-meson decays is essentially characterized by
\cite{Xing07}
\begin{eqnarray}
\frac{\left|{\rm Im}\left(V^*_{ub} V^{}_{cb}\right)\right|}
{|V^*_{ud} V^{}_{cd}|} \simeq \frac{\left|{\rm Im}\left(V^*_{ub} V^{}_{cb}\right)
\right|}{|V^*_{us} V^{}_{cs}|} \simeq A^2 \lambda^4 \eta \simeq
6.3\times 10^{-4} \; ,
\end{eqnarray}
where $\lambda \simeq 0.224$, $A \simeq 0.836$ and $\eta \sim 0.355$ are
the Wolfenstein parameters \cite{PDG}. In other words, it is the smallest inner angle
of all the six CKM unitarity triangles \cite{Luo2010},
\begin{eqnarray}
\phi^{}_{\rm charm} \equiv \arg\left(-\frac{V^*_{ud} V^{}_{cd}}{V^*_{us}
V^{}_{cs}}\right) \simeq A^2 \lambda^4 \eta \simeq 6.3 \times 10^{-4} \; ,
\end{eqnarray}
that sets an upper bound on the weak-interaction parts of charmed CP violation.
That is why the strength of CP violation in the charm sector is at
most of ${\cal O}(10^{-3})$ in the standard model even if there exist significant
final-state interactions.
\begin{figure*}[h]
\centering
\includegraphics[width=7cm]{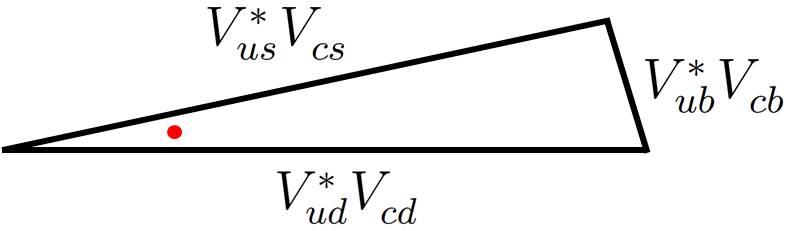}
\vspace{-0.3cm} \caption{The {\it charmed} unitarity triangle of the CKM matrix
defined by the orthogonality relation $V^*_{ud} V^{}_{cd} + V^*_{us} V^{}_{cs} +
V^*_{ub} V^{}_{cb} = 0$ in the complex plane.}
\end{figure*}

\framebox{\bf 2} ~ The above expectation is consistent with the first observation
of direct CP violation in combined $D^0 \to K^+ K^-$ and $D^0 \to \pi^+ \pi^-$ decay
modes, as recently reported by the LHCb Collaboration \cite{LHCb}. The explicit
experimental result is
\begin{eqnarray}
\Delta {\cal A}^{}_{\rm CP} \equiv {\cal A}^{}_{\rm CP} (K^+ K^-) -
{\cal A}^{}_{\rm CP} (\pi^+ \pi^-) = \left(-15.4 \pm 2.9\right) \times 10^{-4} \; ,
\end{eqnarray}
where ${\cal A}^{}_{\rm CP} (K^+ K^-)$ and ${\cal A}^{}_{\rm CP} (\pi^+ \pi^-)$
can simply be interpreted as the direct CP-violating asymmetries of
$D^0 \to K^+ K^-$ and $D^0 \to \pi^+ \pi^-$ decays because both the
$D^0$-$\bar{D}^0$ mixing effect and the indirect CP-violating asymmetries
are found to be negligibly small in this measurement. In this case we just make use of the
definitions
\begin{eqnarray}
{\cal A}^{}_{\rm CP} (K^+ K^-) \hspace{-0.2cm} & = & \hspace{-0.2cm}
\frac{\Gamma(D^0 \to K^+ K^-) - \Gamma(\bar{D}^0 \to K^+ K^-)}
{\Gamma(D^0 \to K^+ K^-) + \Gamma(\bar{D}^0 \to K^+ K^-)} \; , \hspace{0.7cm}
\nonumber \\
{\cal A}^{}_{\rm CP} (\pi^+ \pi^-) \hspace{-0.2cm} & = & \hspace{-0.2cm}
\frac{\Gamma(D^0 \to \pi^+ \pi^-) - \Gamma(\bar{D}^0 \to \pi^+ \pi^-)}
{\Gamma(D^0 \to \pi^+ \pi^-) + \Gamma(\bar{D}^0 \to \pi^+ \pi^-)} \; ,
\end{eqnarray}
and neglect the effects of $D^0$-$\bar{D}^0$ mixing and indirect CP
violation as a fairly reasonable approximation. Then the question is how to separately
determine or estimate the values of ${\cal A}^{}_{\rm CP} (K^+ K^-)$ and
${\cal A}^{}_{\rm CP} (\pi^+ \pi^-)$ from their difference given
in Eq. (3).
\begin{figure*}[h]
\centering
\includegraphics[width=13cm]{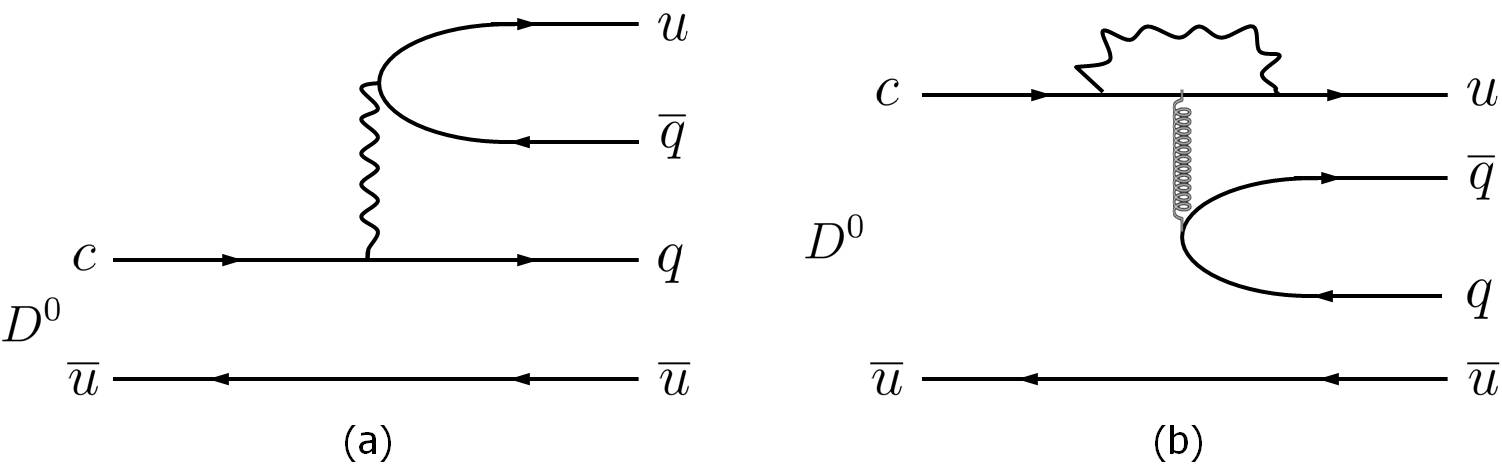}
\vspace{-0.3cm} \caption{The tree and penguin Feynman diagrams for $D^0 \to
K^+ K^-$ and $\pi^+ \pi^-$ decays with $q = d$ and $s$, respectively. Hence there
is a U-spin symmetry between these two decay modes --- a kind of invariance under
the $d \leftrightarrow s$ interchange.}
\end{figure*}


\framebox{\bf 3} ~ Assuming that direct CP violation arises from the interference between
tree and one-loop (penguin) amplitudes of $D^0 \to K^+ K^-$ or $D^0 \to \pi^+ \pi^-$
decay as illustrated in Fig. 2,
we find that ${\cal A}^{}_{\rm CP} (K^+ K^-) = -
{\cal A}^{}_{\rm CP} (\pi^+ \pi^-)$ holds in the limit of U-spin symmetry
(i.e., $d \leftrightarrow s$ interchange symmetry) for the strong-interaction
parts of these two decays
\footnote{U-spin is an SU(2) subgroup of flavor SU(3) group, under which a
pair of down ($d$) and strange ($s$) quarks forms a doublet, analogous to
the isospin symmetry of up ($u$) and down ($d$) quarks. Under this symmetry
$d$ and $s$ quarks are expected to couple equally to gluons at short distances
in all the quark diagrams, and thus the breaking of $d \leftrightarrow s$
interchange symmetry mainly occurs at the hadron level and its effect is
measured by the relevant decay constants and form factors \cite{Lipkin}.}.
To see this point clearly, let us write out their decay amplitudes in a universal
way as follows:
\begin{eqnarray}
A(D^0 \to K^+ K^-) \hspace{-0.2cm} & = & \hspace{-0.2cm}
T^{}_s \left(V^{}_{cs} V^*_{us}\right) \exp\left({\rm i} \delta^{}_s\right)
+ P^{}_s \left(V^{}_{cb} V^*_{ub}\right) \exp\left({\rm i} \delta^{\prime}_s\right)
\; , \hspace{0.7cm}
\nonumber \\
A(D^0 \to \pi^+ \pi^-) \hspace{-0.2cm} & = & \hspace{-0.2cm}
T^{}_d \left(V^{}_{cd} V^*_{ud}\right) \exp\left({\rm i} \delta^{}_d\right)
+ P^{}_d \left(V^{}_{cb} V^*_{ub}\right) \exp\left({\rm i} \delta^{\prime}_d\right)
\; ,
\end{eqnarray}
where $T^{}_q$ and $P^{}_q$ (for $q = d, s$) are real and positive,
$\delta^{}_q$ and $\delta^{\prime}_q$ (for $q = d, s$) stand respectively for
the strong phases of tree and penguin amplitudes, and only the dominant bottom-quark
contribution to the penguin loop is taken into account as a reasonable approximation.
If the penguin-diagram contribution is neglected and the Wolfenstein phase convention
\cite{Wolfenstein} for the CKM matrix is adopted, one will arrive at
$A(D^0 \to K^+ K^-) \simeq - A(D^0 \to \pi^+ \pi^-)$ under U-spin symmetry
\cite{Gronau,Grinstein,Italy} because the latter assures $T^{}_s = T^{}_d$,
$\delta^{}_s = \delta^{}_d$
and $\delta^{\prime}_s = \delta^{\prime}_d$ to hold. Since $K^+ K^-$ and
$\pi^+ \pi^-$ are CP-even eigenstates, it is straightforward to have
\begin{eqnarray}
A(\bar{D}^0 \to K^+ K^-) \hspace{-0.2cm} & = & \hspace{-0.2cm}
T^{}_s \left(V^{*}_{cs} V^{}_{us}\right) \exp\left({\rm i} \delta^{}_s\right)
+ P^{}_s \left(V^{*}_{cb} V^{}_{ub}\right) \exp\left({\rm i} \delta^{\prime}_s\right)
\; , \hspace{0.7cm}
\nonumber \\
A(\bar{D}^0 \to \pi^+ \pi^-) \hspace{-0.2cm} & = & \hspace{-0.2cm}
T^{}_d \left(V^{*}_{cd} V^{}_{ud}\right) \exp\left({\rm i} \delta^{}_d\right)
+ P^{}_d \left(V^{*}_{cb} V^{}_{ub}\right) \exp\left({\rm i} \delta^{\prime}_d\right)
\; .
\end{eqnarray}
Using Eqs. (5) and (6) to calculate the direct CP-violating asymmetries defined
in Eq. (4), we immediately obtain
\begin{eqnarray}
{\cal A}^{}_{\rm CP} (K^+ K^-) \hspace{-0.2cm} & \simeq & \hspace{-0.2cm}
+ 2 A^2 \lambda^4 \eta \frac{P^{}_s}{T^{}_s} \sin\left(\delta^{}_s - \delta^\prime_s
\right) \; , \hspace{0.7cm}
\nonumber \\
{\cal A}^{}_{\rm CP} (\pi^+ \pi^-) \hspace{-0.2cm} & \simeq & \hspace{-0.2cm}
- 2 A^2 \lambda^4 \eta \frac{P^{}_d}{T^{}_d} \sin\left(\delta^{}_d - \delta^\prime_d
\right) \; ,
\end{eqnarray}
where $P^{}_q$ is expected to be comparable with $T^{}_q$
(for $q = d, s$) in magnitude. A combination of Eqs. (3) and (7) yields
\begin{eqnarray}
\frac{P^{}_s}{T^{}_s} \sin\left(\delta^{\prime}_s - \delta^{}_s \right)
+ \frac{P^{}_d}{T^{}_d} \sin\left(\delta^{\prime}_d - \delta^{}_d \right)
\simeq 1.2 \; .
\end{eqnarray}
Given U-spin symmetry, we are left with $P^{}_s/T^{}_s = P^{}_d/T^{}_d$,
$\delta^{}_s = \delta^{}_d$ and $\delta^{\prime}_s = \delta^{\prime}_d$, and thus
\begin{eqnarray}
{\cal A}^{}_{\rm CP} (K^+ K^-) \hspace{-0.2cm} & \simeq & \hspace{-0.2cm}
+\frac{1}{2} \Delta {\cal A}^{}_{\rm CP} \simeq \left(-7.7 \pm 1.5\right)
\times 10^{-4} \; , \hspace{0.7cm}
\nonumber \\
{\cal A}^{}_{\rm CP} (\pi^+ \pi^-) \hspace{-0.2cm} & \simeq & \hspace{-0.2cm}
-\frac{1}{2} \Delta {\cal A}^{}_{\rm CP} \simeq \left(+7.7 \pm 1.5\right)
\times 10^{-4} \; ,
\end{eqnarray}
together with $\left(P^{}_q/T^{}_q\right)
\sin\left(\delta^{\prime}_q - \delta^{}_q \right) \simeq 0.61$ for $q = d$ and $s$.

In view of current experimental data on the branching fractions of $D^0 \to \pi^+ \pi^-$
and $K^+ K^-$ decay modes \cite{PDG},
\begin{eqnarray}
{\cal B}(D^0 \to K^+ K^-) \hspace{-0.2cm} & = & \hspace{-0.2cm}
\left(3.97 \pm 0.07\right) \times 10^{-3} \; ,
\nonumber \\
{\cal B}(D^0 \to \pi^+ \pi^-) \hspace{-0.2cm} & = & \hspace{-0.2cm}
\left(1.407 \pm 0.025\right) \times 10^{-3} \; , \hspace{0.7cm}
\end{eqnarray}
U-spin symmetry is apparently broken. The ratio
$\sqrt{{\cal B}(D^0 \to K^+ K^-)/{\cal B}(D^0 \to \pi^+ \pi^-)}
\simeq 1.68$ cannot be explained unless the relevant phase-space
factors, decay constants and form factors of these two channels 
are all taken into account.
For example, the branching ratios and CP-violating asymmetries
of $D^0 \to \pi^+ \pi^-$ and $K^+ K^-$ decays have recently been recalculated 
by Cheng and Chiang in Ref. \cite{Cheng} and by Li, L$\rm\ddot{u}$ and Yu 
in Ref. \cite{Li} with the consideration of SU(3) symmetry
breaking effects and final-state interactions. Their results 
are essentially consistent with our simpler U-spin
estimates of ${\cal A}^{}_{\rm CP} (K^+ K^-)$ 
and ${\cal A}^{}_{\rm CP} (\pi^+ \pi^-)$ made in Eq. (9). We therefore
expect that the latter should be close to the true values. 

\vspace{0.3cm}

\framebox{\bf 4} ~ Instead of trying to estimate the magnitudes of $T^{}_q$ and
$P^{}_q$ which involve quite a lot of hadronic (nonperturbative) uncertainties,
we proceed to estimate the branching fractions of the CP-forbidden transition
$e^+ e^- \to D^0\bar{D}^0 \to \left(K^+ K^-\right)^{}_D
\left(\pi^+ \pi^-\right)^{}_D$ on the $\psi(3770)$ resonance
with the help of Eqs. (9) and (10). On this resonance the $D^0\bar{D}^0$
pair with odd CP can be coherently produced, and thus its transition into
the CP-even state $\left(K^+ K^-\right)^{}_D \left(\pi^+ \pi^-\right)^{}_D$
is CP-forbidden unless CP is violating. 
Here CP violation is measured by a nonzero rate
rather than an asymmetry between a decay mode and its CP-conjugate progress,
and hence it is of particular interest both theoretically and experimentally.

A generic formula for the branching fraction of such a CP-forbidden transition
has been calculated in Ref. \cite{Xing97}:
\begin{eqnarray}
{\cal B}(f^{}_1, f^{}_2) \hspace{-0.2cm} & = & \hspace{-0.2cm}
{\cal B}(D^0 \to f^{}_1) \cdot {\cal B}(D^0 \to f^{}_2) \cdot
\left(1 + |\lambda^{}_{f^{}_1}|^2\right) \cdot
\left(1 + |\lambda^{}_{f^{}_2}|^2\right) \cdot \left|\frac{p}{q}\right|^2
\nonumber \\
\hspace{-0.2cm} & & \hspace{-0.2cm}
\times \left[\frac{1}{1 - y^2_D} \left(1 - {\cal W}^{}_{f^{}_{1}} {\cal W}^{}_{f^{}_{2}}
\right) - \frac{1}{1 + x^2_D} \left({\cal U}^{}_{f^{}_{1}} {\cal U}^{}_{f^{}_{2}}
+ {\cal V}^{}_{f^{}_{1}} {\cal V}^{}_{f^{}_{2}}\right) \right] \; ,
\end{eqnarray}
where $f^{}_{1}$ and $f^{}_{2}$ are two CP eigenstates with the same CP parity,
$p$ and $q$ denote the complex $D^0$-$\bar{D}^0$ mixing parameters, 
$x^{}_D \equiv \Delta m/\Gamma$ and $y^{}_D \equiv \Delta \Gamma/(2\Gamma)$
stand respectively for the mass and width differences between the two mass 
eigenstates of $D^0$ and $\bar{D}^0$ mesons, and 
\begin{eqnarray}
{\cal U}^{}_{f} \equiv \frac{1 - |\lambda^{}_f|^2}{1 + |\lambda^{}_f|^2} \; ,
\quad {\cal V}^{}_{f} \equiv \frac{-2 {\rm Im} \lambda^{}_f}
{1 + |\lambda^{}_f|^2} \; , \quad
{\cal W}^{}_{f} \equiv \frac{2 {\rm Re} \lambda^{}_f}
{1 + |\lambda^{}_f|^2} \; 
\end{eqnarray}
with $\lambda^{}_f \equiv  (q/p) \cdot A(\bar{D}^0 \to f)/A(D^0 \to f)$. 
Given the fact that both CP violation in $D^0$-$\bar{D}^0$ mixing and indirect 
CP violation from the interplay of decay and $D^0$-$\bar{D}^0$ mixing are 
negligible in $D^0 \to K^+ K^-$ and $D^0 \to \pi^+ \pi^-$ channels \cite{LHCb}, 
which imply $|p/q| \simeq 1$ and 
${\cal V}^{}_{K^+K^-} \simeq {\cal V}^{}_{\pi^+\pi^-} \simeq 0$,
it is convenient for us to simplify Eq. (11) by neglecting the small
$x^2_D$ and $y^2_D$ terms. We arrive at the following formula which only 
contains the direct CP-violating effects:
\begin{eqnarray}
{\cal B}(K^+K^-, \pi^+\pi^-) \hspace{-0.2cm} & \approx & \hspace{-0.2cm}
2 {\cal B}(D^0 \to K^+ K^-) \cdot {\cal B}(D^0 \to \pi^+ \pi^-)
\left|\frac{A(\bar{D}^0 \to K^+ K^-)}{A(D^0 \to K^+ K^-)} -
\frac{A(\bar{D}^0 \to \pi^+ \pi^-)}{A(D^0 \to \pi^+ \pi^-)}\right|^2
\nonumber \\
\hspace{-0.2cm} & \approx & \hspace{-0.2cm}
8 A^4 \lambda^8 \eta^2 {\cal B}(D^0 \to K^+ K^-) \cdot {\cal B}(D^0 \to \pi^+ \pi^-)
\left|\frac{P^{}_s}{T^{}_s} e^{{\rm i}\left(\delta^\prime_s - \delta^{}_s
\right)} + \frac{P^{}_d}{T^{}_d} e^{{\rm i}\left(\delta^\prime_d - \delta^{}_d
\right)} \right|^2
\nonumber \\
\hspace{-0.2cm} & \approx & \hspace{-0.2cm}
32 A^4 \lambda^8 \eta^2 {\cal B}(D^0 \to K^+ K^-) \cdot {\cal B}(D^0 \to \pi^+ \pi^-)
\left|\frac{P^{}_q}{T^{}_q}\right|^2
\nonumber \\
\hspace{-0.2cm} & \approx & \hspace{-0.2cm}
7.1 \times 10^{-11} \left|\frac{P^{}_q}{T^{}_q}\right|^2 \; ,
\end{eqnarray}
where U-spin symmetry has finally been used. One can see that this branching fraction
is at most of ${\cal O}(10^{-10})$ if $P^{}_q$ is comparable with $T^{}_q$ (for
$q = d, s$) in magnitude. Therefore, to see a single event of this kind of CP-forbidden
transition requires at least $10^{10}$ $D^0\bar{D}^0$ pairs on the $\psi(3770)$
resonance for a perfect detection efficiency. A high-luminosity super-$\tau$-charm
factory might be able to do this job in the future. However, to discover such a
tiny CP-forbidden transition at the $5\sigma$ level, at lest $10^{12}$ coherent
$D^0\bar{D}^0$ pairs are needed
\footnote{The author would like to thank H.B. Li for pointing out this challenge.}.

\vspace{0.3cm}

In summary, the recent LHCb discovery of direct CP violation in combined 
$D^0 \to K^+ K^-$ and $D^0 \to \pi^+ \pi^-$ decay modes opens an exciting window to
systematically study CP violation in the charm sector with the help of the ongoing 
and upcoming heavy flavor factories. In this connection we have made a simple
U-spin prediction for the CP-forbidden transition 
$e^+ e^- \to D^0\bar{D}^0 \to \left(K^+ K^-\right)^{}_D
\left(\pi^+ \pi^-\right)^{}_D$ on the $\psi(3770)$ resonance and found its 
branching fraction to be of ${\cal O}(10^{-10})$ or smaller. Although it is 
extremely challenging to observe such a suppressed signal of CP violation, 
the latter deserves our special attention and penetrating search because 
it is simply a rate rather than a conventional CP-violating asymmetry.  

\vspace{0.3cm}

The author would like to thank H.Y. Cheng, H.B. Li and S. Zhou for very timely
and helpful discussions.
This work was supported in part by the National Natural Science Foundation of China
under Grant No. 11835013, and the Ministry of Science and Technology of China under
Contract No. 2015CB856701.

\end{document}